\documentclass[twocolumn]{aa} 
\usepackage{natbib}
\usepackage{amsmath}
\usepackage{graphicx}
\usepackage{txfonts}
\usepackage{textcomp}

\makeatletter
\def\@biblabel#1{}
\makeatother
\newcommand{\mjup}{M${\rm _J}$\,}
\newcommand{\msun}{M$_\odot$\,}
\newcommand{\mstar}{M$_\star$}
\newcommand{\ms}{m\,s$^{-1}$\,}
\newcommand{\teff}{T$_{{\rm eff}}$\,}
\renewcommand{\cite}{\citealp}

\begin{document}

   \title{Study of the impact of the post-MS evolution of the host star on the
orbits of close-in planets.\thanks{Based on observations collected at La Silla - Paranal Observatory under
programs ID's 085.C-0557, 087.C.0476, 089.C-0524 and 090.C-0345.}}

   \subtitle{II. A giant planet in a close-in orbit around the RGB star HIP\,63242}

   \author{M. I. Jones \inst{1,2}
           \and J. S. Jenkins \inst{1}
           \and P. Rojo \inst{1}
           \and C. H. F. Melo \inst{2} 
           \and Paz Bluhm \inst{3}}


   \institute{Departamento de Astronom\'ia, Universidad de Chile, Camino El Observatorio 1515, Las Condes, Santiago, Chile \\\email{mjones@das.uchile.cl}
         \and European Southern Observatory, Casilla 19001, Santiago, Chile
         \and Departamento de Astronom\'ia, Universidad de Concepción, Casilla 160-C, Concepción, Chile }

   \date{}

 
  \abstract
{More than 40 planets have been found around giant stars, revealing a lack of systems orbiting interior to $\sim$ 0.6 AU. This observational fact contrasts with the planetary population around solar-type stars and has been 
interpreted as the result of the orbital evolution of planets due to the interaction with the host star and/or because of 
a different formation/migration scenario of planets around more massive stars.}
{We are conducting a radial velocity study of a sample of 166 giant stars aimed at studying the population of close-in planets
orbiting post-main sequence stars.}
{We have computed precision radial velocities from multi-epoch spectroscopic data, in order to search for planets around
giant stars.}
{In this paper we present the discovery of a massive planet around the intermediate-mass giant star HIP\,63242. The best
keplerian fit to the data lead to an orbital distance of 0.57 AU, an eccentricity of 0.23 and a projected
mass of 9.2 \mjup. HIP\,63242\,b is the innermost planet detected around any intermediate-mass giant star
and also the first planet detected in our survey.}
   {}


   \maketitle
%

\section{Introduction}

So far, more than 800 exoplanets have been detected\footnote{As of March, 2013. Source: http://exoplanet.eu}, most
of them by the radial velocity (RV) technique. The detection
of planets by this method is strongly biased to solar-like host
stars, having low rotational velocity and low levels of stellar activity
(e.g. Jenkins et al. \cite{JEN13}).
Fast rotation broadens the spectral lines, preventing us
from computing precision RV variations, whereas stellar activity and
spots produce spectral line assymmetries, which might mimic the
doppler shift induced by a substellar companion (e.g. Queloz
et al. \cite{QUE01}; Huelamo et al. \cite{HUE08}). In addition, very low mass
stars are too cool and present strong molecular bands in their
spectra, making the computation of precision RV’s more difficult, so
generally only the slowest rotators are targeted (see Jenkins et al. \cite{JEN09}; 
Barnes et al. \cite{BAR12}). 
On the other hand, main-sequence (MS) stars more massive than $\sim$ 1.3
\msun
(corresponding to spectral types earlier than $\sim$ F5) are too
hot and rotate fast, thus leading to an optical spectrum dominated
by few and broad absorption lines. However, after the MS,
early type stars become cooler and rotate slower than their MS
progenitors (e.g. Schrijver \& Pols \cite{SCH93}; Rutten \& Pylyser \cite{RUT98}),
and hence present many narrow absorption lines in their optical
spectra. Also, even though they exhibit a higher level of activity
than solar-type stars, giants with B-V color $<$ 1.2 are quite
stable, and show a stellar jitter at the $\sim$ 20 \ms level (Sato et
al. \cite{SAT05}; Hekker et al. \cite{HEK06}; Jones et al. \cite{JON13}). Therefore, evolved stars present
an ideal case where the RV technique can be applied to search for planets
orbiting intermediate-mass stars (1.3 $\lesssim$ \mstar/\msun $\lesssim$ 3.0) and to
study the post-MS star-planet interactions. \newline
\begin{figure}
\centering
\includegraphics[width=6.8cm,height=10cm,angle=-90]{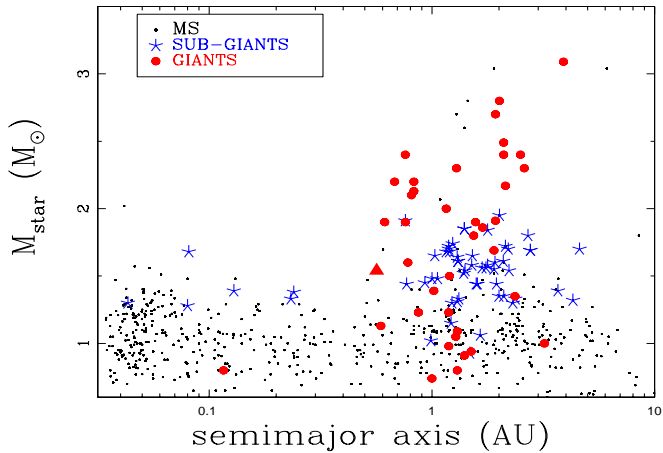}
\caption{Semimajor-axis distribution for planets around evolved stars.
The blue stars and red filled circles correspond to sub-giant and giant
host stars, respectively. The red triangle correspond to HIP\,63242 b. For
comparison, planets around MS stars are also plotted (small black dots).\label{semi_Mstar_1.3Msun}}
\end{figure}
To date, $\sim$ 100 exoplanets have been detected around post-MS
stars (including subgiants), revealing different orbital properties
when compared to the planetary population orbiting FGK
dwarfs. Figure \ref{semi_Mstar_1.3Msun} shows the semimajor axis distribution versus
the mass of the host stars for all of the known planets. The
small black dots, blue stars and red filled circles correspond to
MS, sub-giants and giants host stars, respectively. The red filled
triangle shows the position of HIP\,63242 b. From Figure \ref{semi_Mstar_1.3Msun} it is
evident that there is a lack of close-in orbiting planets (a $\lesssim$ 0.6
AU) around giant stars\footnote{There is only one known planet around giant stars interior to
0.6 AU. The planet is in a 16.2 days period orbit around a 0.8 \msun Horizontal
Branch star (Setiawan et al. \cite{SET10}). However, there is no available parallax
for the host star, and thus its mass and evolutionary status are quite
uncertain.}, whereas there are many short period
planets around MS stars. This
observational result suggests that close-in planets are destroyed
by the large envelope of the host star during the red giant phase.
This idea was predicted theoretically to be caused by the strong
tidal interaction between the planet and the stellar envelope. As
a result, planets orbiting interior to a given distance spiral inward
and are subsequently engulfed by the host star (e.g. Siess
\& Livio \cite{SIE99}; Sato et al. \cite{SAT08}; Villaver \& Livio \cite{VIL09}; Kunitomo
et al. \cite{KUN11}). However, the planetary population around subgiant
stars shows a similar trend, i.e., a deficit of planets orbiting interior
to $\sim$ 0.6 AU. Since subgiant stars still have relatively small
radii, the tidal effect is not expected to significantly affect the
planetary orbits, meaning that stellar evolution cannot be
solely responsible for this observational result (Johnson et al.
\cite{JOH07}). In fact, Bowler et al. \cite{BOW10} showed that the period distribution
of planets around intermediate-mass stars (all of them
detected around subgiants) is different than the population of
planets around FGK dwarfs, at the 4 $\sigma$ level. In particular they
found that planets around intermediate-mass stars present systematically
larger semimajor axis, compared to planets orbiting
low-mass stars. This result might explain in part the planet desert
observed in Figure \ref{semi_Mstar_1.3Msun}, but does not explain the lack of planets
around giant stars with $\sim$ 1.0 - 1.5 \msun\footnote{This result is also attributted 
to a target selection bias in giant stars RV surveys}. \newline  \indent
In this paper we present the detection of a massive giant planet
around HIP\,63242, a nearby G8 giant star. Based on the best keplerian
fit, the minimum mass of HIP\,63242 b is 9.2 \mjup with an
orbital period of 124.6 days, corresponding to a semimajor axis
of a = 0.57 AU. This is the closest planet detected around a first
ascending red giant branch (RGB) star, and the second closest
around a giant star, after HIP\,13044 (Setiawan et al. \cite{SET10}).

\begin{figure}
\centering
\includegraphics[width=6.5cm,height=9cm,angle=-90]{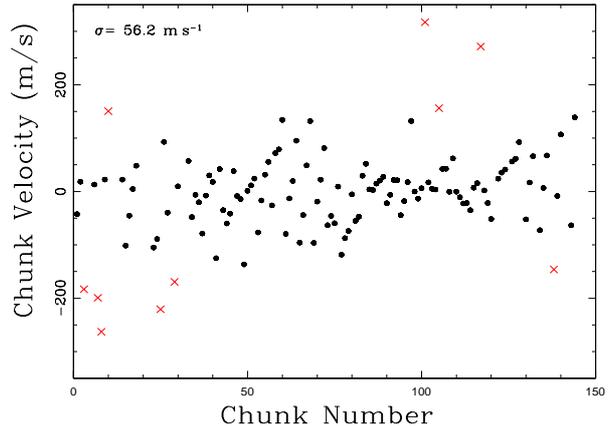}
\caption{Radial velocities computed to 144 different chunks, from
a single observation of $\tau$ Ceti. The solid black dots are the chunks velocities,
while the red crosses correspond to the rejected velocities. The
standard deviation is 56.2 \ms , corresponding to an error in the mean
velocity of 5.3 \ms. \label{chunks}}
\end{figure}


\section{Observations and data reduction}

The data were obatained using FEROS (Kaufer et al. \cite{KAU99}),
mounted on the 2.2m telescope, at La Silla Observatory. The
typical exposure time for the spectra was 210 seconds, leading
to a S/N $>$ 100. The extraction of the FEROS spectra was done 
with the ESO Data Reduction System (DRS), which is available for 
FEROS users. The DRS performs a bias substraction,
flat fielding, orders tracing and extraction. In addition, the scattered
light is substracted. The wavelength calibration was computed
using 4 calibration lamps (one ThAr + one ThArNe, instead of the 12 standard calibrations) having different
exposure times and intensities, which allows coverage of all of the
spectral range ($\sim$ 3500 -9200 Å). The typical RMS in the wavelength
solution is $\sim$ 0.005 \AA. Finally, the wavelength calibration
is applied to the observed spectra, which are extracted order by
order. Additionally, the reduction pipeline applies a barycentric
correction to the extracted spectra, but this option was disabled
because it retrieves the coordinates of the star that are recorded in
the header, which are not accurate enough. Instead, this correction
was computed separately, and then is applied to the reduced
data, as discussed in the next section.

\subsection{Radial velocity calculation}

\begin{figure}
\centering
\includegraphics[width=6cm,angle=270]{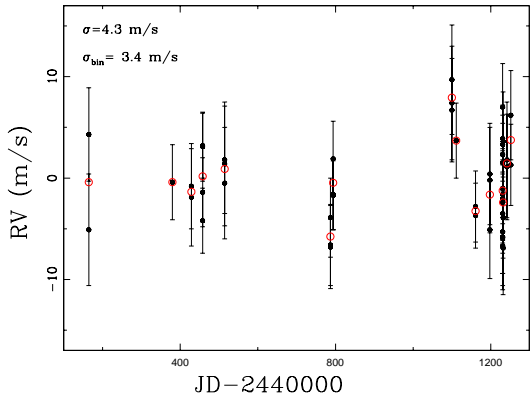}
\caption{Three years observations of the radial velocity standard star $\tau$ Ceti. 
The black dots correspond to the RV's measured from individual FEROS spectra. The RMS is 4.3 \ms. The 
red open circles represent the binned RV's for individual nights. This time the RMS drops to 3.4 \ms.}
\label{tauceti}
\end{figure}

The RV’s for each individual spectrum were measured in the
following manner. Firstly, the doppler shift was computed by
applying a Cross Correlation (Tonry \& Davis \cite{TON79}) between the
stellar spectrum and its corresponding template (high S/N spectrum
of the same star). For this purpose we used the IRAF task
RV/fxcor (Fitzpatrick \cite{FIT93}). This method was applied to $\sim$ 50
Å chunks (corresponding to $\sim$ 1700 pixels), leading to a total
of 144 different RV’s per observation. Then, for each dataset,
the mean velocity was computed, rejecting in an iterative way
every point lying more than 2.5 sigma from the mean, which typically corresponds
to 20 \% of them. It is worth mentioning that since all of
the orders were included, cutting only 100 pixels at the edges,
many chunks lead to very deviating velocities mainly either due
to low S/N (specially toward the blue) or because of the presence
of telluric lines (in the red part of the spectrum). \newline 
Figure \ref{chunks} shows the chunk velocities from one spectrum of $\tau$ 
Ceti\footnote{$\tau$ Ceti is a kwown stable star at the few \ms level. 
However, Tuomi et al. \cite{TUO12}, have shown that it hosts a planetary system} 
The black dots are the non-deviant velocities and the red crosses
are those rejected by the procedure just described (some of them
are out of the plotting region). The standard deviation of the chunk velocities
is 56.2 \ms , which corresponds to an error in the mean\footnote{The error in the mean is 
given by: $\sigma$/$\sqrt(n_c)$, where $\sigma$ is the standard deviation of the chunks 
velocities and n$_c$ is the number of non-rejected chunks used in the analysis} of just
5.3 \ms .
The second step consists of a similar procedure, but this time the
cross correlation is computed between the simultaneous calibration
lamp (sky fiber) and one of the lamps that was used for the
wavelength calibration of that night (i.e., corresponding to the
night zero point), having a similar exposure time to the simultaneous
calibration lamp. This procedure is neccessary to substract
the nighlty drift, produced mainly by small variations in the refraction
index of air in the spectrograph during the night, which at first
order translates into a linear RV shift as large as $\sim$ 150 \ms . It
is worth to mention that no second order correction was applied,
like the RV shift between the two fibers, which is typically $\sim$ 
2-3 \ms (Setiawan et al. 2000). 
Finally, the radial velocity for each epoch is computed by:
\begin{equation}
{\rm RV = RV_{ob,tem} + RV_{drift} + BC}  
\end{equation}
where the first and second terms correspond to the RV computed
for the object with its corresponding template and to the
nightly drift, as explained above. The third term corresponds to
the barycentric correction, which is computed using the mean  
time of the observation and using the actual coordinates of the
star at that time, which are slightly different to the ones recorded
in the image header (typically up to $\sim$ 1-2 arcminutes). This is
quite important, since the error in the header coordinates translates
into a RV uncertainty as big as $\sim$ 5-10 \ms. 
Finally, we tested the long-term precision of FEROS using 52 spectra of
$\tau$ Ceti, taken in 17 different nights (one spectrum was used as template) during the last three 
years. The resulting RV's are shown in Figure \ref{tauceti} (black dots). The measured RMS is 4.3 \ms. 
In addition, we binned the RV datapoints for individual nights, in order to average out the main stellar
oscillations modes\footnote{The typical exposure time of the $\tau$ Ceti spectra is $\sim$ 10-30 seconds.}  
(e.g. O'toole et al. \cite{OTO08}). The binned radial velocities (red open circles) lead to a RMS of only
3.4 \ms. This result shows the huge potential of FEROS for high precision RV studies of bright stars.


\section{HIP\,63242\,b: {\rm the closest planet around and
intermediate-mass giant star}}

\begin{figure}
\centering
\includegraphics[width=8.0cm,height=10cm,angle=-90]{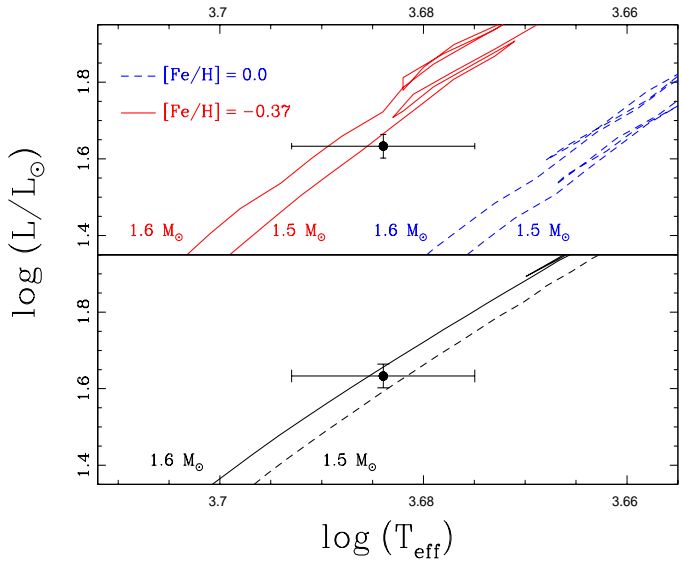}
\caption{Upper panel: Position of HIP\,63242 in the HR diagram. The four closest evolutionary
tracks from Salasnich et al. \citep{SAL00} are overplotted. Lower panel: Same as for the upper panel,
but this time using stellar tracks from the Yonsei-Yale database. Both models were interpolated to 
[Fe/H] = -0.31.  \label{new_HIP63242_position}}
\end{figure}

According to the Hipparcos catalogue, HIP\,63242 is a G8III star
with V=6.87, B-V=1.03 and a parallax of $\pi$=7.42 $\pm$ 0.49, which
correspond to a distance of 135 pc. 
As for the rest of the stars in our sample, we derived \teff, [Fe/H] and log\,$g$ for HIP\,63242, using
the equivalent width of iron lines (Fe\,{\sc i} and Fe\,{\sc ii}), by imposing excitation and 
ionization equilibrium. In order to do this we used the MOOG 
\footnote{http://www.as.utexas.edu./$\sim$chris/ moog.html} code (Sneden \cite{SNE73}) along with the Kurucz 
(\cite{KUR93}) atmosphere models. For a more detailed description see Jones et al. (\cite{JON11}). Additionally,
we computed the luminosity of HIP\,63242 using the bolometric corrections given in Alonso et al. (\cite{ALO99}) and
the 3-D extinction maps of Arenou et al. (\cite{ARE92}).
We compared the resulting \teff, [Fe/H] and stellar luminosity, which are listed in Table 1, with Salasnich et al. \citep{SAL00} evolutionary 
models. We derived a mass of 1.54 \msun for HIP\,63242, using a linear interpolation method, as described in Jones 
et al. \cite{JON11}.
Figure \ref{new_HIP63242_position} shows the position of
HIP\,63242 in the HR diagram and the closest evolutionary tracks 
from Salasnich et al. \citep{SAL00} (upper panel). This star is clearly ascending the
RGB, since no HB model cross its position in the HR diagram.  
For comparison,  
Figure \ref{new_HIP63242_position} (lower panel) also shows two evolutionary isomass tracks taken from 
the Yonsei-Yale evolutionary models (Demarque et al. \cite{DEM04}) with solar-scaled metal abundances 
([$\alpha$/Fe]=0.0).
Both tracks were interpolated to [Fe/H]=-0.31, using the interpolator included with the evolutionary models
\footnote{www.astro.yale.edu/demarque/yystar.html}. As can be seen, the mass for 
HIP\,63242 derived using both sets of models is almost identical.
\begin{table}
\centering
\caption{Stellar properties of HIP\,63242\label{tab1}}
\begin{tabular}{l c}
\hline\hline
Parameter &  Value  \\
\hline
B\,-\,V (mag)    &  1.02 $\pm$ 0.02 \\
V (mag)          &  6.86 $\pm$ 0.01 \\
$\pi$ (mas)      &  7.42 $\pm$ 0.49 \\
\teff (K)        &  4830 $\pm$ 100 \\
log\,g (cm\,s$^{-2}$) &  2.53 $\pm$ 0.2 \\
${\rm [Fe/H]}$ (dex)    &    -0.31 $\pm$ 0.09 \\
L (L$_\odot$)   &  42.7  $\pm$ 0.08 \\
Mass (\msun)    &  1.54 $\pm$ 0.05 \\
v\,sin$i$ (km\,s$^{-1}$) & 3.7 $\pm$ 0.1 \\
\hline\hline
\end{tabular}
\end{table}

\begin{figure}
\centering
\includegraphics[width=7.0cm,height=10cm,angle=-90]{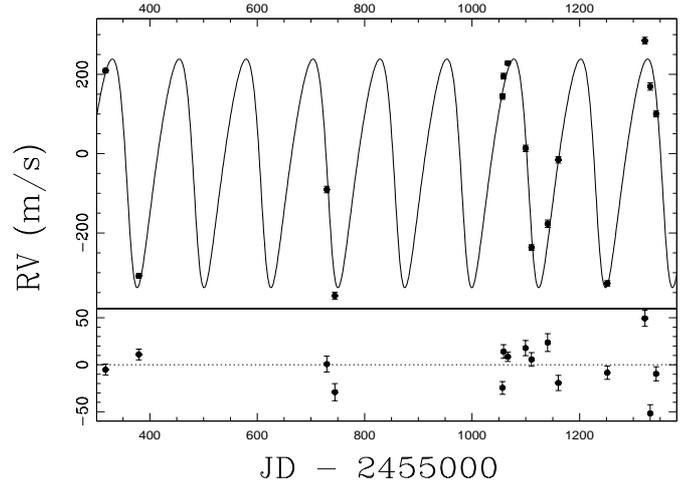}
\caption{RV curve for HIP\,63242 (black dots). The best keplerian fit is
overplotted (black solid line). \label{HIP63242}}
\end{figure}

During the last three years, 16 spectra (including the template) of HIP\,63242
were taken with FEROS. Its RV curve is shown in Figure \ref{HIP63242}. 
The resulting velocities are also listed in Table \ref{HIP63242_RV} (available in the electronic
version). The error bars are $\sim$ 5\,-\,8 \ms, therefore
are smaller than the symbol sizes. The best keplerian fit \footnote{The keplerian solution 
was computed using the Systemic Console (Meschiari et al. \cite{MES09})} is
overplotted (solid black curve). It can be seen that there is a
strong RV signal present in the data. The orbital parameters of
the planet are listed in Table 2. The RMS of the fit is 23.7 \ms,
which is mainly explained by stellar jitter. However, it is also possible that the presence of 
a second planet in the system produces a larger scatter from the single planet fit. Unfortunately, there are
not enough observations yet to test this hypothesis. 
\begin{table}
\centering
\caption{Orbital parameters of HIP\,63242\,b \label{tab2}}
\begin{tabular}{l c}
\hline\hline
Parameter      &    Value  \\
\hline
P (days)       &   124.6   \\
K (\ms)        &   287.5   \\ 
a (AU)         &   0.565   \\
e              &   0.23    \\  
$\omega$ (deg) &   118.2   \\
T$_0$ (JD)     &   2455376.2 \\
M$_p$\,sin$i$  &   (M$_J$) 9.18  \\
\hline\hline
\end{tabular}
\end{table}

\section{Photometric and Line Profile Analysis}

Intrinsic stellar phenomena, such as spots, magnetic activity or
stellar oscillations, are known to produce periodic RV signal,that can mimic the doppler shift induced by a substellar companion
(e.g. Queloz et al. \cite{QUE01} Huelamo et al. \cite{HUE08}; Figueira
et al. \cite{FIG10}). We performed three standard tests aimed at determine whether 
this is the case for HIP\,63242.
First, we analyzed the Hipparcos photometric data, which consists of a
total of 142 H$_p$ filter observations, taken between JD 2447869
and 2449013. The photometric data show a small dispersion of
0.009 mag, which cannot be responsible for the observed large
RV variations. In fact, according Hatzes (2002), a spot covering
5\% of the stellar surface ($\Delta$m $\sim$ 0.06 mag) induces a RV variation
below 100 \ms, on a star having a projected rotational
velocity similar to HIP\,63242 (v$_{rot}$ = 3.6 k\ms). Also, no significant
periodic signal is observed in the Hipparcos photometry.
Hence, rotational modulation can be discarded as the responsible
mechanism for the observed RV signal. Also, we did a bisector
analysis (Toner \& Gray 1988), aimed at detecting asymmetries
in the line profiles caused by intrinsic stellar phenomena.
\begin{figure}
\centering
\includegraphics[width=7.0cm,height=9cm,angle=-90]{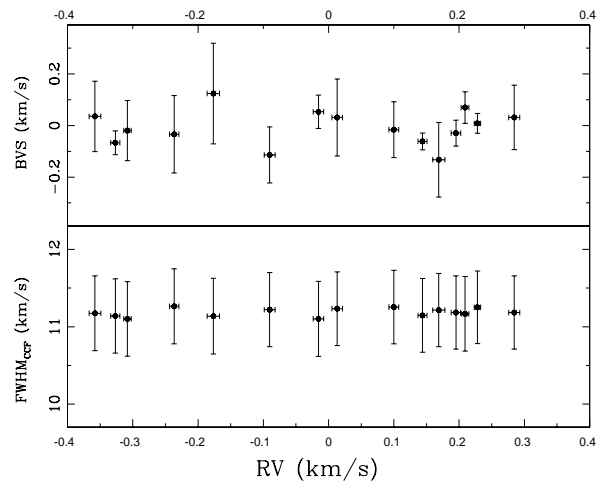}
\caption{Bisector velocity span (upper panel) and FWHM of the CCF (lower panel) 
against the RV’s measured for HIP\,63242. In both cases the CCF’s were computed for 11
different orders, covering the wavelength range between $\sim$ 5000\,-\,6000 \AA.
The errorbars correspond to the uncertainty in the mean. \label{fig5}}
\end{figure}
\begin{figure}
\centering
\includegraphics[width=6.5cm,height=8cm,angle=-90]{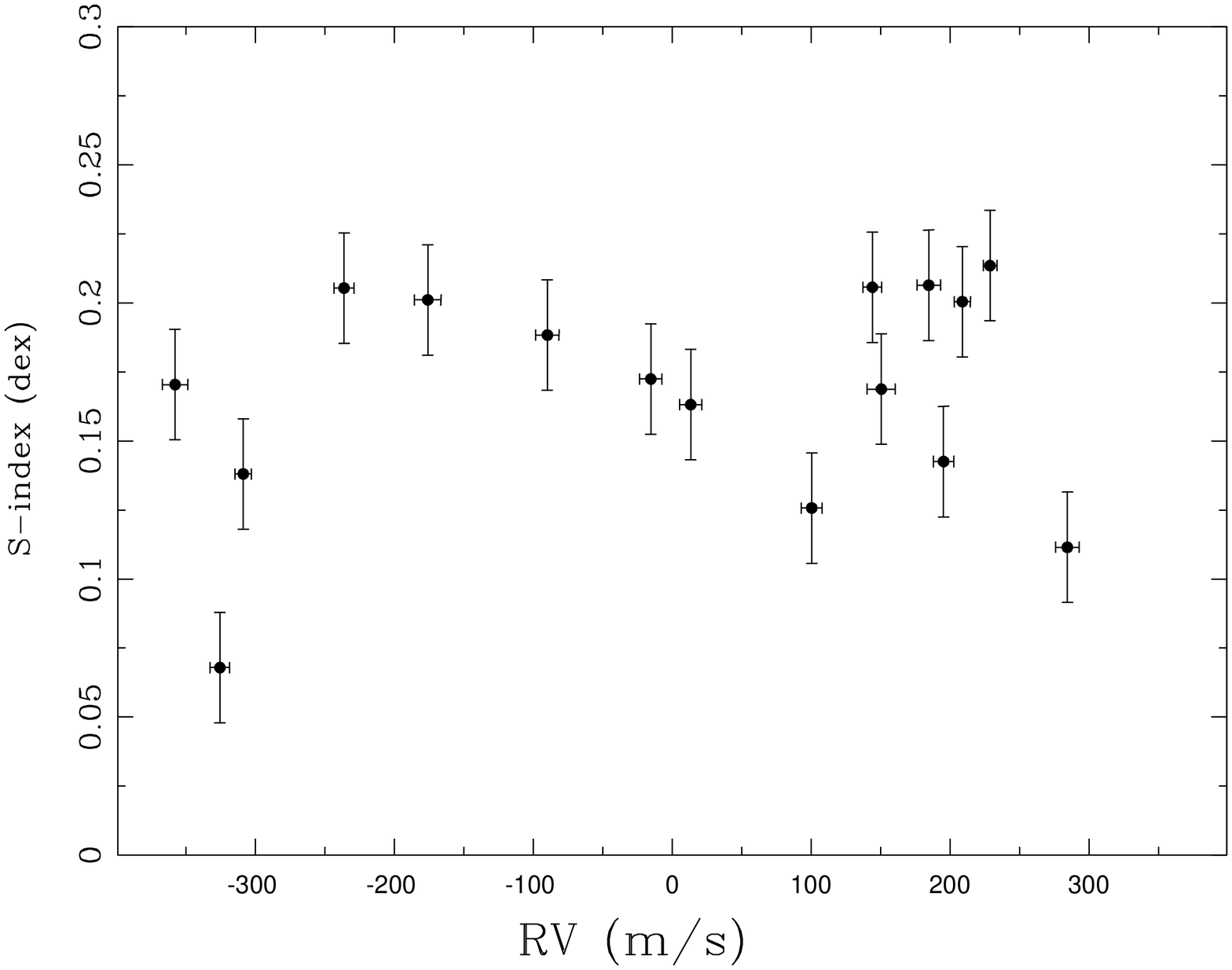}
\caption{S-index variation as described in Jenkins et al. (\cite{JEN08,JEN11}) against the measured radial
velocities for HIP\,63242. \label{HIP63242_S_index}}
\end{figure}
Figure \ref{fig5} (upper panel) shows the bisector velocity span
(BVS), which corresponds to the velocity difference between the
bottom and the top of the CCF\footnote{We computed the average BVS from 11 different orders.
The errorbars correspond to the error in the mean.}, versus
the observed radial velocities for HIP\,63242. Clearly no obvious correlation between
both quantities is present. Also, in the lower panel the width of the CCF as a function of the measured, 
RV’s is plotted. Once again
there is no correlation between both quantities. In both
cases, the RMS around the mean is comparable to the errorbars.
Finally, Figure \ref{HIP63242_S_index} shows the S-index variations, computed in a similar fashion as described in 
Jenkins et al. \citep{JEN08,JEN11}, against the observed radial velocities. No
correlation seems to be present. Based on these
stellar activity diagnostics, we can conclude that the most likely
explanation for the RV signal observed in HIP63242 is due to
the presence of a substellar companion.
\onltab{
\begin{table}
\centering
\caption{Radial velocity measurements of HIP\,63242 \label{HIP63242_RV}}
\begin{tabular}{ccc}
\hline\hline
JD\,-        & RV & error  \\
2455000   &  (\ms)   &   (\ms) \\
\hline
 317.5802 &  209.1 & 5.9 \\
 379.4784 & -308.2 & 5.8 \\
 729.4990 &  -90.3 & 8.4 \\
 744.4786 & -358.0 & 9.1 \\
1056.5393 &  143.9 & 6.8 \\
1058.6287 &  195.2 & 7.3 \\
1066.5553 &  228.1 & 4.9 \\
1099.4790 &   13.2 & 8.1 \\
1110.5131 & -236.6 & 7.3 \\
1140.5548 & -176.5 & 9.5 \\
1160.5382 &  -15.5 & 8.0 \\
1251.8552 & -326.8 & 7.0 \\
1321.7227 &  284.6 & 8.5 \\
1331.7437 &  168.9 & 9.3 \\
1342.6957 &  100.0 & 7.5 \\
\hline\hline
\end{tabular}
\end{table}
}

\section{Summary and conclusions}

We computed precision radial velocities using FEROS spectra of
the giant star HIP\,63242, which have revealed a large periodic
signal. We developed a radial velocity computation method that
leads to long-term RV precision of $\sim$ 3-4 \ms, which is
much better than was previously obtained with FEROS data.
To determine whether these variations are related to intrinsic
stellar phenomena (rotational modulation, stellar pulsation
or magnetic-related activity), we performed a detailed photometric,
line profile and Ca II lines emission analysis. We found
no correlation with the RV variations, meaning that the
observed radial velocity signal is likely attributed to an extrinsic
mechanism. \newline \indent
According to the best keplerian fit, and assuming a mass for the host star of 1.54 \msun,
we derived a semimajor axis of 0.57 AU for
HIP\,63242 b, which correspond to the innermost planet detected
around a RGB star.
The detection of
these kind of planets is very important because allow us to better
understand what is the effect of the stellar evolution (after the
MS) in the orbital properties of planets. In addition, even though
close-in planets around intermediate mass stars are rare, more of
them can be expected to be detected in the coming years, allowing
us to disentangle the effect of the stellar mass from the stellar
evolution in their orbits.

\begin{acknowledgements}
We acknowledge the referee, John A. Johnson for his very useful comments.
M.J. and P.R. acknowledge financial support from Fondecyt through grant \#1120299.
M.J. also acknowledges financial support from ALMA-Conicyt grant \#31080027 and
from BASAL PFB-06.
J.J. acknowledges funding by Fondecyt through grant 3110004 and
partial support from CATA (PB06, Conicyt), the GEMINI-CONICYT FUND and
from the Comit\'e Mixto ESO-GOBIERNO DE CHILE. 
\end{acknowledgements}

\end{document}